\documentclass[journal]{IEEEtran}

\IEEEoverridecommandlockouts

\usepackage[utf8]{inputenc} 

\usepackage[T1]{fontenc}
\usepackage{url}
\usepackage{ifthen}
\usepackage{cite}
\usepackage{overpic}
\usepackage{balance}
\usepackage[cmex10]{amsmath} 

\usepackage{textcase}
\usepackage[tablename=Table]{caption}
\usepackage{blindtext}
\usepackage{floatrow}
\usepackage{soul}
\usepackage{changes}
\usepackage{gensymb}
\usepackage{cite}
\usepackage{amsmath,amssymb,amsfonts}
\usepackage{textcomp}
\usepackage{graphicx}
\usepackage{amssymb}
\usepackage{amsfonts}
\usepackage{amsmath}
\usepackage{epsfig}
\usepackage{epigraph}
\usepackage{color}
\usepackage{fancybox}
\usepackage{textcomp}
\usepackage{multirow}
\usepackage{setspace}
\usepackage{psfrag}
\usepackage{booktabs}
\usepackage{float}
\usepackage[caption = false]{subfig}
\usepackage{algorithm}
\usepackage{algpseudocode}
\usepackage{siunitx}
\usepackage{mathtools, nccmath, bigints, amsfonts}
\usepackage{array}
\usepackage{stfloats}
\newcolumntype{L}{>{\centering\arraybackslash}m{3cm}}

\usepackage{amsthm}
\theoremstyle{definition}
\usepackage{mathrsfs}
\usepackage{mdframed}
\usepackage{tcolorbox}

\newfloatcommand{capbtabbox}{table}[][0.4\textwidth]
\definecolor{matlabgreen}{rgb}{0.0, 0.5, 0.0}

\usepackage{pict2e} 

\theoremstyle{plain}

\usepackage{algpseudocode} 

\newcommand{\vect}[1]{\mathbf{#1}}

\def\Htran{\mbox{\tiny $\mathrm{H}$}}
\def\Ttran{\mbox{\tiny $\mathrm{T}$}}

\def\NUE{N}
\def\DUE{\Delta}

    \def\Complex{{\rm\rule[.23ex]{.03em}{1.1ex}\kern-.3em{C}}}

    \newcommand{\be}{\begin{equation}} \newcommand{\ee}{\end{equation}}
    \newcommand{\bea}{\begin{eqnarray}} \newcommand{\eea}{\end{eqnarray}}
    \newcommand{\benum}{\begin{enumerate}} \newcommand{\eenum}{\end{enumerate}}

\def\Htran{\mbox{\tiny $\mathrm{H}$}}
\def\Ttran{\mbox{\tiny $\mathrm{T}$}}

\begin{document}

\title{Pre-Optimized Irregular Arrays versus Moveable Antennas in Multi-User MIMO Systems}

\author{Amna~Irshad,~Alva~Kosasih,~Vitaly~Petrov, and Emil~Bj{\"o}rnson,~\IEEEmembership{Fellow,~IEEE}\\

\thanks{A. Irshad, V. Petrov and E. Bj{\"o}rnson are with the Division of Communication Systems,  KTH Royal Institute of Technology, SE-100 44 Stockholm, Sweden. Email: \{amnai,vitalyp,emilbjo\}@kth.se.}
\thanks{A. Kosasih was with KTH and is now with Nokia, Finland. Email: alva.kosasih@nokia.com }
\thanks{The research was supported by the Grant 2022-04222 from the Swedish Research Council.
The computations were enabled by resources provided by the National Academic Infrastructure for Supercomputing in Sweden (NAISS), partially funded by the Swedish Research Council (Grant no. 2022-06725).
}

}

\maketitle

\begin{abstract}
Massive multiple-input multiple-output (MIMO) systems exploit the spatial diversity achieved with an array of many antennas to perform spatial multiplexing of many users. Similar performance can be achieved using fewer antennas if movable antenna (MA) elements are used instead. MA-enabled arrays can dynamically change the antenna locations, mechanically or electrically, to achieve maximum spatial diversity for the current propagation conditions. However, optimizing the antenna locations for each channel realization is computationally excessive, requires channel knowledge for all conceivable locations, and requires rapid antenna movements, thus making real-time implementation cumbersome. To overcome these challenges, we propose a pre-optimized irregular array (PIA) concept, where the antenna locations at the base station are optimized a priori for a given coverage area.
The objective is to maximize the average sum rate and we take a particle swarm optimization approach to solve it. Simulation results show that PIA achieves performance comparable to MA-enabled arrays while outperforming traditional uniform arrays.
Hence, PIA offers a fixed yet efficient array deployment approach without the complexities associated with MA-enabled arrays.

\end{abstract}


\section{Introduction}
\label{sec:intro}
The adoption of massive MIMO (multiple-input multiple-output) technology with $32$-$64$ antennas per base station (BS) was groundbreaking for 5G, enhancing both spectral and energy efficiency compared to 4G~\cite{massivemimobook}. Hence, the number of antennas per BS is envisioned to continue growing in 6G and beyond, for example, in the form of ultra-massive MIMO at terahertz frequencies~\cite{Faisal2020VTMag} and gigantic MIMO in the upper mid-band~\cite{2024_Bjornson_Arxiv}. These evolutions will lead to uniform planar arrays (UPAs) with hundreds or even thousands of antennas per BS, which is associated with significant increases in hardware cost and computational complexity due to the numerous parallel radio frequency (RF) chains~\cite{Mietzner2009Survey}. This raises the question: \emph{Can we design non-uniform arrays that achieve similar performance using substantially fewer antennas?}

In multi-user MIMO scenarios, the BS antenna array should capture enough spatial diversity to make the users' channel matrices nearly orthogonal. This can be achieved in low-scattering propagation environments using a sparse array configuration at the BS. 
Thanks to the increased spacing between the antennas, the spatial correlation among them decreases, leading to more variability among the channel coefficients observed over the array (i.e., more spatial diversity) compared to using a conventional half-wavelength-spaced array~\cite{Do2021a}. Even if sparse arrays achieve very narrow beams, they also suffer from grating lobes that might hit other users and anyway cause strong interference \cite{bookEmil}---this issue cannot be fully mitigated using an array with fixed antenna locations. 

To fully exploit spatial diversity, the concept of movable antennas (MAs) was proposed in \cite{BJORNSON20193} and recently evaluated in~\cite{10753482,cheng2023sumrate, xiao2023multiuser} (among others). The principle idea is to use a flexible antenna array where the individual elements can be mechanically or electrically ``moved'' to a different location (within certain bounds), e.g., using a flexible cable/rail and a controller such as a stepper motor~\cite{movgeneral}.
The main feature of \emph{MA-enabled arrays} is that the array geometry can be tailored to the current needs of the system~\cite{10753482}, such as creating exact orthogonality between the current users' channel matrices.

An MA-enabled array must be equipped with real-time optimization functionalities.
In~\cite{cheng2023sumrate,xiao2023multiuser}, an optimization framework utilizing particle swarm optimization (PSO) algorithm was proposed to move the antennas to maximize the sum rate for the current users. Using that framework, the MA-enabled array demonstrates significant performance gains over fixed uniform arrays. A recent review of the topic is found in \cite{10753482}.

There are three inherent practical challenges with using MAs in a real system. First and foremost, it requires a \emph{dynamic} MA---a movable antenna that can change its configuration when already in use---which is a step beyond current antenna technology. Second, the run-time of the optimization algorithm must be a fraction of the channel coherence time (i.e., at the sub-millisecond level), which is a major challenge since PSO is a computationally intense optimization technique.
Finally, running the optimization algorithm for every channel realization implies that the BS side must have perfect knowledge of the channel conditions for all conceivable antenna locations, which is possible in free-space propagation with known user positions, but rarely the case in real network deployments. Hence, while the use of MA-enabled arrays can lead to major performance gains in theory, the three above-mentioned issues are so severe that they might not be practically realizable. Furthermore, to the best of our knowledge, existing studies on optimization for MA-enabled arrays focus exclusively on single-antenna multi-user MIMO systems. However, 5G user equipments (UEs) are already equipped with multiple antennas, which are used to enhance spectral efficiency, and these must be considered in any practical MA-enabled system.

In this paper, we propose addressing the problems with MA-enabled arrays by developing an antenna array optimization framework, where the BS antenna locations are optimized offline and only once (i.e., before its deployment on the site). The resulting array is sparse, irregular, and tailored to the coverage area, but the antenna configuration remains fixed during system operation, so there is no need to dynamically reconfigure it on-site. Our optimization framework is based on the well-established PSO approach, but we propose a novel objective function---to maximize the expected sum rate across all possible UE locations in the given coverage area---and tailor the PSO algorithm for this goal.

The main contributions of this paper are:
\begin{itemize}
     \item We address the high complexity problems related to MA-enabled arrays by \textbf{formulating a novel objective function---referred to as \emph{pre-optimized irregular array (PIA)}}---where the \emph{average} sum rate is maximized, rather than the \emph{instantaneous} sum rate used in state-of-the-art approaches. As a result, different from MA approaches, no real-time optimization of the array is required by PIA, thus, notably reducing the implementation complexity. We show numerically that PIA still achieves nearly the same performance as MA-enabled arrays, particularly when there is a surplus of BS antennas, and significantly outperforms conventional and sparse UPA.
     \item  Our optimization framework is developed for a system where the \textbf{UEs are equipped with multiple antennas}. This extension relaxes a key simplistic assumption in the existing works on MA-enabled arrays: the focus on single-antenna multi-user MIMO systems. This generalization is \emph{non-trivial}, as it requires incorporating advanced signal processing techniques, such as block diagonalization, among other additions.
\end{itemize}

Section II presents the system model, followed by Section III and Section IV, which discuss the problem formulation and our proposed solution, respectively. Section V presents the numerical evaluation of the proposed solution, \emph{PIA}, in comparison with benchmarks such as MA-enabled arrays.

\section{System Model}
\label{sec:system_model}
\begin{figure}[t!]
\includegraphics[width=0.95\textwidth]{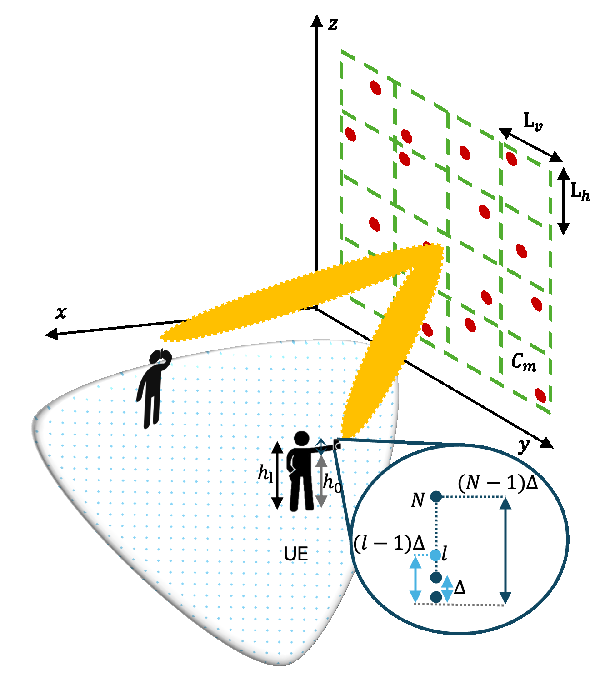}         
\caption{Illustration of BS with $M$ MAs serving $K$ UEs, each is equipped with $\NUE$ antennas, in the  deployment region.}
\vspace{-5mm}
\label{system model}
\end{figure}

We consider a communication system with a BS equipped with a sparse irregular array, as illustrated in Fig.~\ref{system model}.
 The array consists of $M$ antennas that are
deployed in the $yz$-plane. The coordinate of the $m$-th BS antenna is denoted by $\vect{t}_m=[0,y_m,z_m]^{\Ttran}$, for $m \in \{1,\dots,M\}$. 
The antenna coordinates will be optimized, but for practical reasons, they can only be moved within predefined non-overlapping rectangular areas.
More precisely, we consider $M$ reference positions $(y_{m}^0,z_{m}^0)$ that are arranged as a sparse UPA in the $yz$-plane with $M_{\rm  v}$ rows and $M_{\rm  h}$ columns, where $M=M_{\rm  v}M_{\rm  h}$.
The $m$-th antenna can be moved within a local $2$D region $\mathcal{C}_m$ of size $L_{\rm h}\times L_{\rm v}$, centered around a reference position, such that $\mathcal{C}_m = \{(y, z) \mid y_{m}^0-\frac{L_{\rm h}}{2} \leq y \leq y_{m}^0+\frac{L_{\rm h}}{2}, \, z_{m}^0-\frac{L_{\rm v}}{2} \leq z \leq z_{m}^0+\frac{L_{\rm v}}{2}\}$.

We consider a downlink scenario where the BS transmits to $K$ multi-antenna UEs. Each UE is equipped with $\NUE$ antennas, arranged in a uniform linear array (ULA) along the $z$-axis, with an antenna spacing of $\DUE$.
The position vector for the $\ell$-th antenna of the $i$-th UE is defined as $\vect{r}_{i,\ell}=[r_x^{i},r_y^{i},{h}_{\ell}]^{\Ttran}$, for $i \in \{1,\dots,K\}$ and $\ell \in \{1,\dots,\NUE\}$, where $r_x^{i}$ and $r_y^{i}$ are the $xy$-coordinates of the UE, and ${h}_{\ell}$ is the height of the $\ell$-th antenna. This height is defined as $h_{\ell} = h_{0} + (\ell - 1)\DUE$, where $h_{0}$ is the reference height and $\ell$ is the antenna index.\footnote{The exact UE antenna configuration only affects the numerical results.} To ensure efficient spatial multiplexing with controllable inter-user interference, we assume that $M \geq \NUE K$.\footnote{If there are more user antennas than BS antennas, then user scheduling is normally used to ensure that $M \geq \NUE K$~\cite{uescheduling}.} We further assume perfect channel knowledge to focus on the optimization of the BS antenna locations.

The $n_i$-dimensional data vector intended for the $i$-th UE, $\vect{d}_i \in \mathbb{C}^{n_i}$, is multiplied with a precoding matrix $\vect{W}_i \in \mathbb{C}^{M \times n_i}$ and sent from the BS array. 
The received signal $\mathbf{y}_i \in \mathbb{C}^{\NUE}$  is
 \begin{align}\label{eq_rec_signal}
     \vect{y}_i= \vect{H}_i \vect{W}_i \vect{d}_i +  \sum_{j=1,j\neq i}^K \vect{H}_i \vect{W}_j  \vect{d}_j + \vect{n}_i,
\end{align}
where the channel matrix to the $i$-th UE is denoted by $\vect{H}_i \in \mathbb{C}^{\NUE \times M}$ and $\vect{n}_i \in \mathbb{C}^{\NUE}$ represents the additive complex circular-symmetric Gaussian noise vector with the covariance matrix $\sigma^2 \vect{I}$. 
The second term in \eqref{eq_rec_signal} represents the interference experienced by the $i$-th UE from data transmitted to other UEs.

The multi-UE interference term in \eqref{eq_rec_signal} can be eliminated if the precoder matrices  $\vect{W}_j$ for all UEs are designed such that $\vect{H}_i \vect{W}_j = \vect{0},  \forall i \neq j$, i.e., the signal intended for the $j$-th UE lies in the null space of the channel matrices of all other UEs. This can be achieved using the block diagonalization (BD) technique~\cite{spencer}, which constructs each UE's precoder matrix to ensure zero interference. For the $i$-th UE, the combined channel matrix of all other UEs is denoted by 
\begin{equation}
\overline{\vect{H}}_i = \left[ \vect{H}_1^{\Ttran}, \dots, \vect{H}_{i-1}^{\Ttran}, \vect{H}_{i+1}^{\Ttran}, \dots, \vect{H}_K^{\Ttran} \right]^{\Ttran} \in \mathbb{C}^{\NUE(K-1) \times M}.
\end{equation}
The precoder matrix $\vect{W}_i $ must lie in the null space of $\overline{\vect{H}}_i$ to eliminate inter-UE interference. Let the column of $\vect{\overline{V}}^{(0)}_i$ contain an orthonormal basis for the null space of $\vect{\overline{\vect{H}} }_i$, such that $\overline{\vect{H}}_i \vect{\overline{V}}_i^{(0)} =\vect{0}$.
Once the precoder matrix is restricted to the null space of the interfering UEs' channels, the precoder is designed to maximize the achievable rate for the $i$-th UE. This is done by applying singular value decomposition (SVD) to the effective channel $\vect{H}_i \vect{\overline{V}_i}^{(0)} $, which leads to
\begin{equation}
   \vect{H}_i \vect{\overline{V}}^{(0)}_i=\vect{U}_i
\begin{bmatrix}
\vect{\Sigma}_i & \vect{0}\\ \vect{0} &\vect{0}
\end{bmatrix} \left[\vect{V}^{(1)}_i \vect{V}^{(0)}_i \right]^{\Htran}. 
\end{equation}
The precoder matrix for the $i$-th UE is then constructed as~\cite{spencer}
\begin{equation} \label{eq:BD_precoding}
   \vect{W}_i=  \vect{\overline{V}}^{(0)}_i \vect{{V}}^{(1)}_i \vect{D}_i^{1/2},
\end{equation}
where $\vect{{V}}^{(1)}_i$ contains the $n_i$ singular vectors corresponding to non-zero singular values, $\vect{D}_i \in \mathbb{C}^{n_i \times n_i}$ is a diagonal power allocation matrix, and $(\cdot)^{1/2}$ denotes a matrix square root.

\section{Problem Formulation}
\label{sec:problem}

In this section, we present a novel problem formulation that aims to optimize the BS antenna location to achieve maximal communication performance, using a metric defined below.

For any BD precoding matrix of the kind in \eqref{eq:BD_precoding}, we can compute the rate achieved by the $i$-th UE as
\begin{equation}\label{eq_SumRate}
    R_i=\log_2  \det{ \left( \vect{I}+\frac{1}{\sigma^2} \vect{H}_i\vect{W}_i\vect{W}_i^{\Htran}\vect{H}_i^{\Htran} \right)}.
\end{equation}
For fixed antenna locations, we can maximize the sum rate $R_{\Sigma}= \sum_{i=1}^{K} R_i$ under a total downlink transmit power of $p_{\rm max}$. The optimal power allocation matrices $\vect{D}_1,\ldots,\vect{D}_K$ are given by the water-filling technique~\cite{spencer,bookEmil}, which allocates power between UEs and individual data streams.

Since the UE locations are predetermined and uncontrollable, the only remaining design variable for sum-rate optimization is the positioning of the BS antennas. Each channel matrix $\vect{H}_i$ is actually a function of the BS antenna locations $\vect{t}_m,  \forall m$, which can be expressed as $\vect{H}_i (\vect{t}_1,...,\vect{t}_M)$.
This type of optimization has been considered in the existing literature, for example, in~\cite{cheng2023sumrate,xiao2023multiuser} (among others), where the antenna locations are optimized to maximize the SINR/rate for one set of UEs. When the UEs move or the scheduling changes, the optimal antenna locations must be updated. This requires real-time moveable antennas, which is associated with significant implementation complexity, cost, and latency as optimization must be performed once per channel coherence time.

In contrast, we propose to optimize the antenna locations only once, based on the statistics of the user population and coverage area. Suppose the user locations are generated as a realization $\omega$ from a random variable $\Omega$. The sum rate for these user locations can then be expressed as  $R_{\Sigma}^{(\omega)}(\vect{t}_1,...,\vect{t}_M)$, where we also highlight the dependence on the BS antenna locations. We want to maximize the average sum rate across different realizations of user locations.

We formulate our optimization problem as follows:
\begin{align} \label{optproblem}
\underset{{\vect{t}_1,\ldots, \vect{t}_M}}{\text{maximize}} \quad & \mathbb{E} \left\{ R_{\Sigma}^{(\omega)}(\vect{t}_1,...,\vect{t}_M) \right\}, \\ \label{cona}
\text{subject to} \quad & {\vect{ t}_m} \in \mathcal{C}_m, \quad 1\leq m \leq M, \\ \label{conb}
& {\| \vect{t}_m-\vect{t}_j \|}_2 \geq \lambda/2, \quad 1\leq m \neq j \leq M,  
\end{align}

where the expectation is taken with respect to the random variable $\Omega$ (i.e., the locations of the $K$ UEs). 
The constraint \eqref{cona} allows each antenna element to move within the specified region $\mathcal{C}_m$, while the constraint \eqref{conb} ensures a minimum separation distance  $ \lambda/2$ between any two antenna elements at the BS to avoid practical mutual coupling issues.

We call the optimal solution to \eqref{optproblem} the \emph{pre-optimized irregular array (PIA)}.
The optimization problem is non-convex and no analytical solution can be derived, but we will propose an algorithmic solution in the next section.

\section{Proposed Array Optimization Framework}
\label{sec:proposed_framework}

In this section, we explain our proposed method for optimizing the BS antenna locations to solve the problem formulated in \eqref{optproblem}, thereby finding the PIA.
We leverage the PSO method~\cite{PSObookClerc}, which is a moderate-complexity optimization framework. PSO optimizes the objective function by iteratively evaluating $N_{p}$ candidate solutions, called particles. Each particle represents a set of BS antenna locations (i.e., an antenna array configuration). The search space is defined by the constraints in \eqref{cona} and \eqref{conb}. At each iteration, each particle ($p$) moves around in the search space but keeps track of its own best-known position $ \vect{P}_{\rm best}^p \in \mathbb{R}^{M \times 3}$, $\forall p \in \{1,...,N_{p}\} $ and the swarm particles' overall best-known position $ \vect{G}_{\rm best}\in \mathbb{R}^{M \times 3}$. Note that the $m$-th row of $ \vect{P}_{\rm best}^p$ and $ \vect{G}_{\rm best}$ represents a potential location for the $m$-th antenna for the $p$-th particle. 
 
Initially, the velocity $\vect{V}^p(0) \in \mathbb{R}^{M \times 3}$ and position $\vect{X}^p(0) \in \mathbb{R}^{M \times 3}$ of all particles are randomly initialized within the defined constraints. The algorithm proceeds as follows:

\begin{enumerate}
    \item \textbf{Evaluate the objective function:} Compute the value of the sum-rate objective function for each particle, as described in Algorithm~\ref{APSOalgo}. The initial antenna locations are given by $\vect{X}^p(0)$.
    \item \textbf{Update $ \vect{P}_{\rm best} $ and $ \vect{G}_{\rm best} $:}
    \begin{itemize}
        \item $ \vect{P}_{\rm best}^p $: The position with the highest objective function value found so far for the $p$-th particle.
        \item $  \vect{G}_{\rm best} $: The position with the highest objective function value across all particles.
    \end{itemize}
    \item \textbf{Update velocity and position:}
    \begin{itemize}
        \item Compute the new velocity of each particle as
        \begin{multline}
       \vect{V}^p(t+1) = \upsilon \vect{V}^p(t) + c_1 u_1 \big(\vect{P}_{\rm best}^p - \vect{X}^p(t)\big) \\+ c_2 u_2 \big(\vect{G}_{\rm best} - \vect{X}^p(t)\big),
        \end{multline}
        where $ \upsilon $ is the inertia weight, $ c_1 $ and $ c_2 $ are acceleration coefficients, $  u_1, u_2 \sim \mathcal{U}(0,1) $ are random numbers uniformly distributed in $[0,1]$, and $t$ denotes the iteration inside the PSO algorithm.
        \item Update the position of each particle as
        \begin{equation}
        \vect{X}^p(t+1) = \vect{X}^p(t) + \vect{V}^p(t+1).
        \end{equation}
    \end{itemize}
    \item \textbf{Termination:}
         The algorithm terminates if the maximum number of iterations denoted as $n_{\rm PSO}$ is reached. Otherwise, it returns to Step 1.
\end{enumerate}
Once terminated, the algorithm outputs the final objective value and the final BS antenna locations represented by $ \vect{G}_{\rm best} $. It is important to note that this optimization procedure can be conducted offline, similar to the training process used in deep learning techniques, so computational complexity is not a concern. The optimization can be treated as part of the cell planning, so we first identify the optimal locations and then let an engineer deploy the antennas accordingly. If significant changes occur in the propagation environment, the offline training can be repeated to determine a more suitable array configuration for the new conditions and the antennas can be moved by an engineer or using mechanical methods.

The objective function in \eqref{optproblem} is generally hard to evaluate exactly, but it can be approximated using a sample average as
\begin{equation}
    \mathbb{E} \left\{ R_{\Sigma}^{(\omega)}(\vect{t}_1,...,\vect{t}_M) \right\} \approx \frac{1}{Q} \sum_{\omega \in \mathbb{Q}} R_{\Sigma}^{(\omega)}
\end{equation}
for a set $\mathbb{Q}$ containing $Q$ random realizations from the random variable $\Omega$. Inspired by the stochastic gradient descent approach in deep learning, we consider a new set of  $Q$ random realizations for each iteration of the PSO algorithm so we can keep $Q$ reasonably small without overfitting on those realizations. This way of computing the PSO objective value is summarized in Algorithm~\ref{APSOalgo}.

\begin{algorithm}
\small
\caption{ PSO Objective Function}
\begin{algorithmic}[1]
    \State \textbf{Input:} Antenna locations $\vect{t}_1,\ldots,\vect{t}_M$, the random variable $\Omega$ generating user locations, and number of user realizations $Q$
     \State Initialize $\hat{R}_{\Sigma}=0$
    \For{$q = 1, \dots, Q$}
        \State Randomly distribute $K$ UEs using the random variable $\Omega$
        \State Generate the channel matrices $\vect{H}_i(\vect{t}_1,\ldots,\vect{t}_M)$ for all users
        \State Compute the sum rate $R_{\Sigma} ^{(\omega)}$ and update $\hat{R}_{\Sigma} \gets \hat{R}_{\Sigma} + \frac{1}{Q} R_{\Sigma} ^{(\omega)}$
    \EndFor

    \State \textbf{Return:} $\hat{R}_{\Sigma}$

\end{algorithmic}
\label{APSOalgo}
\end{algorithm}

\vspace{-3mm}
\section{Numerical Results}
\label{sec:results}
We evaluate the communication performance of the proposed PIA by comparing it against existing BS array configurations. In the proposed scheme, each BS antenna can be moved within a local 2D region of size $L_{\rm h}\times L_{\rm v}$, where we set $L_{\rm h}=L_{\rm v}=5 \lambda$ m. The center of the BS array is positioned at a height $h_{\rm BS}$ above the $y$-axis.

We compare the PIA against the following benchmarks:  
\begin{itemize}  
   \item \textbf{MA-enabled array:} The BS is equipped with an MA array, where the antenna locations are optimized using PSO for each realization of the user locations. 
   
    \item \textbf{Uniform Sparse planar array (USPA):} The BS is equipped with a sparse UPA with an antenna spacing of $5\lambda$, occupying the same area as the proposed PIA.

     \item \textbf{Half-wavelength planar array (HwPA):} The BS is equipped with a UPA with half-wavelength antenna spacing in both the horizontal and vertical directions. 
\end{itemize}

For a fair comparison, the total number of BS antennas, $M$, is kept the same for all four configurations. Further, the maximal physical dimensions of the BS are the same for the MA-enabled array, USPA, and the proposed PIA. We consider downlink transmission to $K=6$ dual-antenna users who have a uniformly distributed radial distance $\rho \sim \mathcal{U}[\rho_{\min},\rho_{\max}]$ and uniformly distributed angle $\varphi \sim \mathcal{U}[\varphi_{\min},\varphi_{\max}]$ in the $xy$-plane.
We consider the near-field-compliant free-space line-of-sight channel model from considered~\cite{LOSmodel}. The results are evaluated on independent UE locations, different from those used during optimization, to demonstrate the generalizability of the proposed approach and avoid overfitting to specific UE locations. Key parameters are summarized in Table~\ref{Parameters}. 

\begin{table}[t!]
    \centering
    \caption{Summary of simulation parameters}
    \label{Parameters}
   \begin{tabular}{ p{1.2cm}||p{5.1cm}||p{1cm}}
    \hline
    \textbf{Variable} & \textbf{Description} & \textbf{Default value}\\
    \hline
    \multicolumn{3}{c}{\emph{Radio and antenna parameters}}\\
    \hline
    $\NUE$ & Number of antennas at each UE & 2\\
    $\DUE$ & Antenna spacing at UEs (m)& $1\lambda$  \\
    $h_{\rm BS}$ & Reference height of the BS (m) & $40\lambda$ \\
    $h_{0}$ & Reference height of the UE (m) & $1.25$   \\
    $p_{\max}$ & Transmitted power at the BS (W) & 50  \\
    $\sigma^2$& Noise variance (pW) & $3.98$   \\
    $f_c$ & Carrier frequency (GHz) & 3 \\
    $B$ & Bandwidth (MHz) & $100$   \\
    $\rho_{\min}$ & Radial lower limit (m) & $20\lambda  $  \\
    $\rho_{\max}$ & Radial upper limit (m) & $5000\lambda  $  \\
    $\varphi_{\min}$ & Azimuth lower limit (rad) &  $-\pi/3 $ \\
    $\varphi_{\max}$ & Azimuth upper limit (rad) &  $\pi/3  $  \\
    \hline
    \multicolumn{3}{c}{\emph{Optimization parameters}}\\
    \hline
    $n_{\rm PSO}$&  Number of PSO iterations & 200  \\
    $N_{p}$ & Number of particles in PSO& 150\\
    $w$&PSO inertia weight&0.5\\
    $c_1$&PSO personal coefficient&1.2\\
    $c_2$&PSO global learning coefficient&2\\
    $Q$ & Number of user realizations  & 1000  \\ 
    \hline
    \end{tabular} \vspace{-3mm}
\end{table}

\begin{figure}[b!]
\vspace{-5mm}
\includegraphics[width=0.9\textwidth]{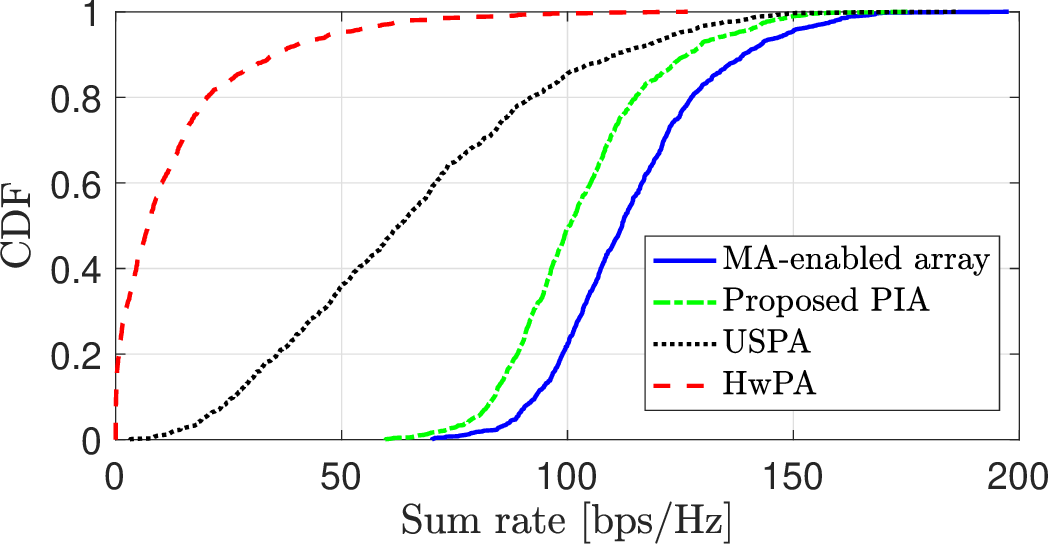}
\vspace{-1mm}
\caption{The  CDF for $M=4^2$ BS antennas.}
\label{fig_case1}
\end{figure}

We first consider the cumulative distribution function (CDF) with respect to the user locations for all the evaluated schemes in Fig.~\ref{fig_case1}, for a BS array with $M=4^2=16$ antennas. The graph demonstrates that the MA-enabled array outperforms the other configurations. This is due to its ability to fine-tune the antenna locations and exploit spatial diversity to nearly orthogonalize the user channels, but it comes at the expense of higher latency, complexity, and cost than having fixed antenna locations.
Our proposed PIA method performs noticeably better than other fixed array configurations, with only a $11\%$ gap in average sum rate to the MA-enabled array. The USPA improves the average sum rate as compared to the compact HwPA since it achieves a narrower beamwidth (at the expense of grating lobes). However, it is not as effective as our PIA method, which is optimized to achieve the maximal average sum rate in the coverage for any fixed array configuration.

\begin{figure}
\centering
{\includegraphics[width=0.9\textwidth]{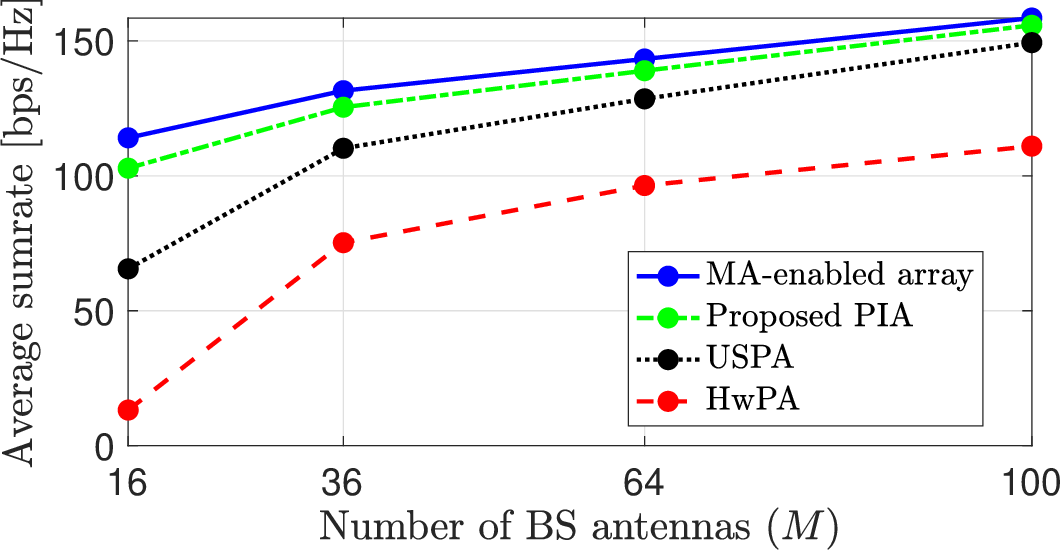}}\label{case2_user6}
\vspace{-1mm}
\caption{The impact of increasing the number of BS antennas on the average sum rate across all considered schemes.}
\label{Var_mean_case2}
\end{figure}

In Fig.~\ref{Var_mean_case2}, we show the average sum rate over $100$ user locations as a function of the number of BS antennas with $M_{\rm  h}=M_{\rm  v} \in \{ 4, 6, 8, 10\}$. We observe that the gap between the MA-enabled array and the proposed PIA method reduces as the number of antennas increases. 
The reason is that user channels always become closer to orthogonal when $M/K$ increases---the phenomenon that underpins the Massive MIMO concept \cite{massivemimobook}. Similar trends can be observed in the case of $K=8$ UEs, but those results are omitted for space limitations. 
These results highlight the benefits of PIA in practical scenarios as we eliminate the need for re-optimization the antenna locations for every channel realization (needed by MA-enabled arrays) while the performance reduction is tiny.

\begin{figure}[t!]
\includegraphics[width=0.9\textwidth]{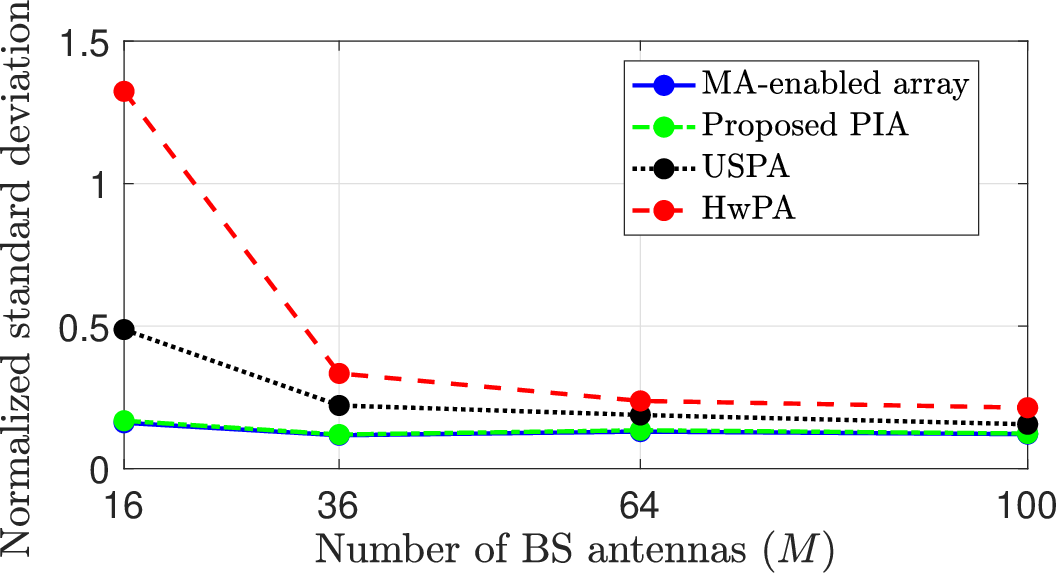}
\vspace{-1mm}
\caption{The variability ratio across BS antennas.}
\label{fig_case3}
\end{figure}

Not only is the gap between the average sum rates shrinking as $M$ increases, but also the variability between different sets of user locations. We demonstrate this in Fig.~\ref{fig_case3} by plotting the normalized standard deviation $\frac{\sqrt{\rm {Var}\{R\}}}{\mathbb{E}\{R\}}$, where ${\rm {Var}\{\cdot\}}$ denotes the variance operator.
Our proposed PIA method exhibits the least variability among the fixed arrays, on par with the MA-enabled array. This result is expected as the objective function in our framework is to maximize the average sum rate in the entire coverage area. Additionally, we observe that as the number of BS antennas increases, all methods converge to a small variability ratio, which is a kind of channel hardening effect.

\vspace{-3mm}
\section{Conclusion}
\label{sec:conclusions}

In this paper, we optimized the BS antenna array geometry for serving multiple multi-antenna UEs. We provided a novel optimization framework to design a pre-optimized irregular array (PIA) tailored to a particular coverage area, in terms of maximizing the average sum rate across different user locations. We showed that a PIA can provide great sum-rate improvements as compared to traditional compact and sparse uniform planar arrays. We also compared PIA with MA-enabled arrays that require optimization for each channel realization.
Remarkably, we demonstrated that a PIA achieves nearly the same performance as an MA-enabled array, particularly when there are more BS antennas than UE antennas. Hence, it is a practically attractive deployment solution since MA-enabled arrays have high implementation complexity, cost, and latency. Notably, the proposed PSO algorithm can be used in any propagation environment with known channel~$\vect{H}$.

\balance
\bibliographystyle{IEEEtran}
\bibliography{IEEEabrv,mybib}

\end{document}